# BioTD: an online database of biotoxins

Gaoang Wang[1,†], Hang Wu[2,†], Yang Liao[1,†], Zhen Chen[1], Qing Zhou[2], Wenxing Wang[2], Yifei Liu[1], Yilin Wang[1], Meijing Wu[2], Ruiqi Xiang[2], Yuntao Yu[1], Xi Zhou[2,*], Feng Zhu[1,*], Zhonghua Liu[2,*], Tingjun Hou[1,*]

[1]College of Pharmaceutical Sciences, Zhejiang University, Hangzhou, Zhejiang 310058, China

[2]The National and Local Joint Engineering Laboratory of Animal Peptide Drug Development, College of Life Sciences, Hunan Normal University, Changsha, Hunan 410081, China

[†]Equivalent authors

[*]Correspondence

**Email**: tingjunhou@zju.edu.cn (Tingjun Hou); liuzh@hunnu.edu.cn (Zhonghua Liu); zhufeng@zju.edu.cn (Feng Zhu); xizh@hunnu.edu.cn (Xi Zhou).

## Abstract

Biotoxins, mainly produced by venomous animals, plants and microorganisms, exhibit high physiological activity and unique effects such as lowering blood pressure and analgesia. A number of venom-derived drugs are already available on the market, with many more candidates currently undergoing clinical and laboratory studies. However, drug design resources related to biotoxins are insufficient, particularly a lack of accurate and extensive activity data. To fulfill this demand, we develop the Biotoxins Database (BioTD). BioTD is the largest open-source database for toxins, offering open access to 14,607 data records (8,185 activity records), covering 8,975 toxins sourced from 5,220 references and patents across over 900 species. The activity data in BioTD is categorized into five groups: Activity, Safety, Kinetics, Hemolysis and other physiological indicators. Moreover, BioTD provides data on 986 mutants, refines the whole sequence and signal peptide sequences of toxins, and annotates disulfide bond information. Given the importance of biotoxins and their associated data, this new database was expected to attract broad interests from diverse research fields in drug discovery. BioTD is freely accessible at http://biotoxin.net/.

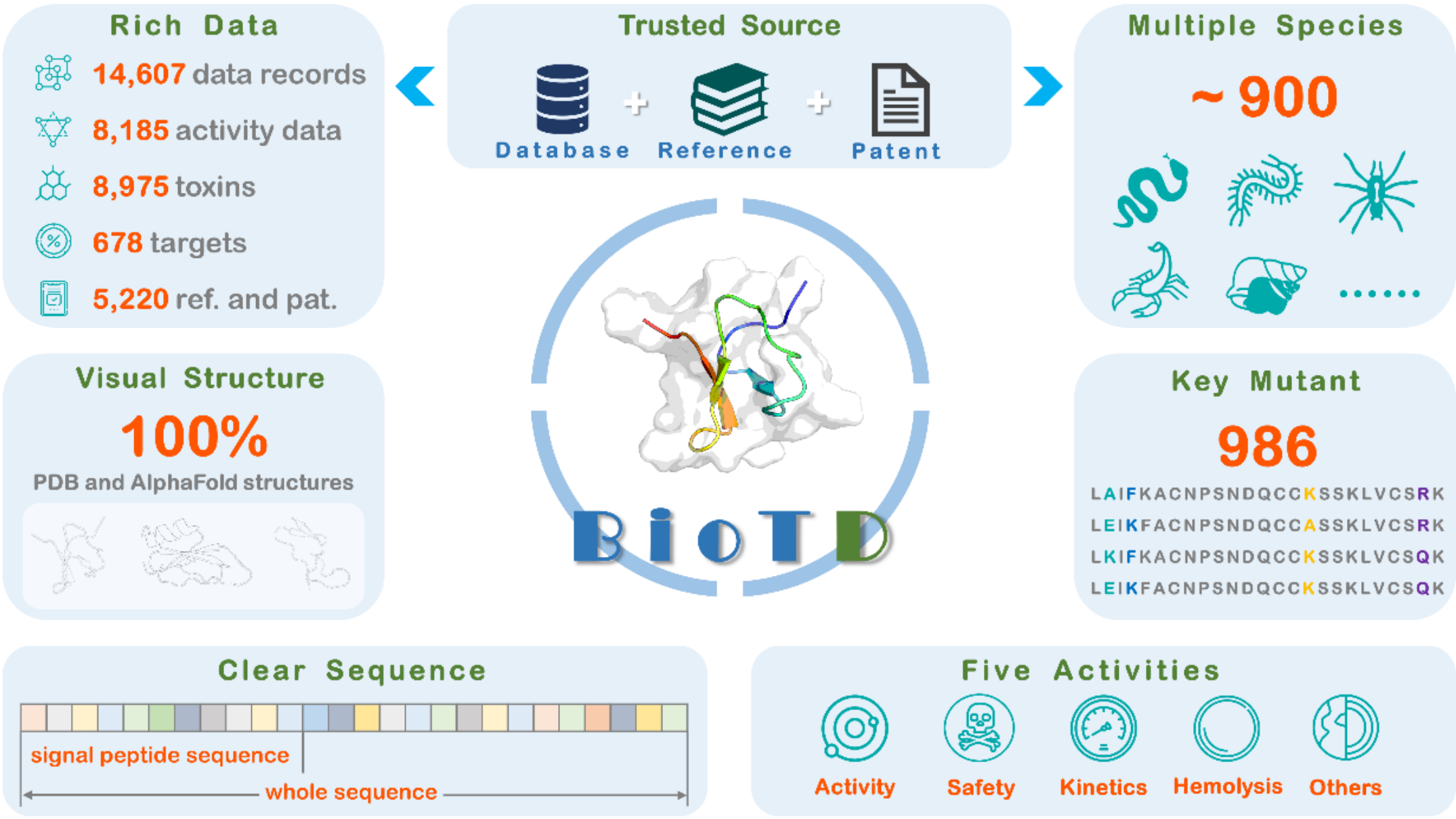

**Graphical Abstract**

# 1 INTRODUCTION

Biotoxins, which primarily consist of peptides and proteins derived from venomous animals, plants, and microbes, are evolved bioactive molecules. These toxins can bind efficiently and sometimes selectively to various protein targets, such as ion channels, receptors, and enzymes, thereby modifying their target function or physiological state. Due to the unique activity and therapeutic effects of the active molecules in venom (e.g., analgesia and blood pressure reduction), biotoxins have been proved as exceptional pharmacological tools for exploring the structures and functions of ion channels or receptors, as well as a crucial resource for innovative drug development. As shown in **Table 1**, a number of venom-derived drugs have been approved by U.S. Food and Drug Administration (FDA), such as Bivalirudin, Captopril, Desirudin, Exenatide, and Ziconotide, and more potential peptide drugs are in preclinical or clinical development[1]. It is worth noting that designing biotoxins-based peptide drugs requires not only basic information such as the sequences and structures of biotoxins but also the activity information of biotoxins to shed light on structure-activity-relationship (SAR) analysis.

To date, several knowledgeable databases such as Uniprot[2], ChEMBL[3], PDB[4], and DrugBank[5] have been developed to provide biotoxins-related information, including peptide sequences, miscellaneous functions, and structure models as part of a broader collection of biological/chemical information. While databases like VenomZone (https://venomzone.expasy.org/) and Kalium[6] offer limited toxin information for specific targets or species, they have attracted considerable attention from the research community. However, none of these databases specifically focuses on collecting comprehensive and systematic data on biotoxins or providing experimentally validated activity information for them. In other words, there is a high demand for a database that comprehensively provides such data.

In our study, we report the Biotoxins Database (BioTD), the largest biotoxins database that provides comprehensive information for toxins from multiple perspectives. First, a systematical literature review is conducted by searching various keywords such as ‘biotoxin’, ‘toxin activity’, ‘toxin mutant’, and ‘toxin species’ in PubMed and public databases, which results in a total of 14,607 data records and 8,975 toxins among more than 900 species, such as snakes, spiders, scorpions, snails, and others, covering almost every type of toxins known. These data come from 5,220 references and patents. Second, given that 986 peptides are mutants modified based on natural peptides, the sequence information of the prototypes and the mutants are both included. Furthermore, BioTD refines the sequence of toxins, distinguishes the whole sequence and the signal peptide sequence, and marks the disulfide bond information, which will help users extract valuable toxin sequence information. Third, based on our review, a total of 8,185 activity information is collected from experimentally verified activity information from 2,122 reviewed literatures. The activity data of BioTD can be divided into five categories: Activity (such as effective concentration/dose, inhibition concentration), Safety (lethal concentration/dose, therapeutic index), Kinetics (such as $K_d$, $K_i$, $K_{act}$, $Tau_{(off)}$), Hemolysis (such as hemolysis rate, minimum coagulant dose, minimum hemagglutination concentration) and other physiological indicators. Finally, all data are seamlessly cross-linked to well-established databases, such as AlphaFold DB[7], UniProt[2], and NCBI Taxonomy[8]. To the best of our knowledge, BioTD is the largest biotoxin knowledgebase that have ever been developed. Due to the importance of the biological activity and structure-related information of biotoxin and the severe lack of such valuable data in existing databases, BioTD is highly expected to attract broad research interests such as medicine, neuroscience, pharmacology, drug discovery, and agriculture. BioTD is freely accessible to all users without any login requirement at: **http://biotoxin.net/**.

**Table 1. A list of worldwide venom-derived drugs that have been approved for clinical use.**

| Drug | Trade name | Species | Disease | Company | First approval | Reference |
|---|---|---|---|---|---|---|
| Batroxobin* | Baquting® | Pit viper (snake) | Perioperative bleeding | Nuokang Biopharma | 2020 | [9] |
| Bivalirudin | Angiomax® | Hirudo medicinalis (medicinal leech) | Coagulation during surgery | Sandoz Inc | 2000 | [10, 11] |
| Captopril | Capoten® | Bothrops jararaca (snake) | Hypertension, congestive heart failure | Bristol-Myers Squibb | 1981 | [12] |
| Desirudin | Iprivask® | Hirudo medicinalis (medicinal leech) | Thrombosis | Bausch Health | 2003 | [13] |
| Enalapril | Vasotec® | Bothrops jararaca (snake) | Hypertension, cardiac failure | Bausch Health | 1985 | [14] |
| Eptifbatide | Integrilin® | Sistrusrus miliarius barbourin (snake) | Thrombosis | Cor Therapeutics, acquired by Millennium Pharmaceuticals | 1998 | [15] |
| Exenatide | Byetta® | Heloderma suspectum | Type2 diabetes mellitus | Amylin | 2005 | [16] |
| | | (lizard) | | Pharmaceuticals | | |
| Lepirudin** | Refludan® | Hirudo medicinalis (medicinal leech) | Heparin-induced thrombocytopenia | Celgene Europe Limited | 1998 | [17] |
| Tirofiban | Aggrestat® | Echis carinatus (snake) | Acute coronary syndrome | Medicure | 1998 | [18] |
| Ziconotide | Prialt® | Conus magus (cone snail) | Treatment of severe and chronic pain | Neurex, acquired by Elan Pharmaceuticals | 2004 | [19] |

*Batroxobin was not currently approved by the FDA and was mainly used in some regions such as China.

**In 2012, lepirudin was discontinued, although this was not due to inefficacy or adverse effects, but rather production issues.

# 2 MATERIALS AND METHODS

## 2.1 Data collection and processing

The data of BioTD were collected from literatures and public databases. The literatures were searched in the PubMed database using the keywords such as 'biotoxin', 'toxin activity', 'toxin mutant', and 'toxin species', etc. We manually collected and checked the important biotoxin information from the literatures, including the sequence, disulfide bond, biological activity, species, target and etc. We also collated the biotoxin and target information in multiple well-known databases, including Uniprot, PDB, DrugBank and AlphaFold DB. It was worth noting that all data in BioTD were supported by corresponding references and patents to ensure the reliability, and information without any reference/patent support was not collected.

To make the access and analysis of BioTD data convenient for all readers, the collected raw data were carefully cleaned up and then systematically standardized. Firstly, almost all of biotoxins (except mutants), targets, species, structures in BioTD were cross-linked to established databases. Secondly, BioTD provided the visual protein structures of all toxins (mutants display the wide-type structures) derived from the NCBI and PDB database or precited by AlphaFold software to provide users with the visual interface. Moreover, BioTD refined the sequence of toxins by using the NovoPro tool[20], distinguished the whole sequence and the signal peptide sequence, and marked the disulfide bond information, which would help users extract valuable toxin sequence information. In brief, we had done our best to provide users with high-quality biotoxin data.

## 2.2 Online database implementation

The website was written in PHP language, combined with the MySQL database solution to provide a system that could support the needs of users to manage and manipulate data. The text search was based on the full name, abbreviation, synonym and gene name of biotoxins and targets. Biotoxin 3D structure display could be visualized by 3Dmol.js tool or obtained online via Uniport ID. Users could obtain biotoxin information similar to their sequence in the database by uploading the peptide FASTA file, and the sequence similarity search function was provided by BLAST-2.6.0+. All technical operations of BioTD were in line with the specifications.

## 3 RESULTS AND DISCUSSIONS

### 3.1 Statistics in BioTD

BioTD has a comprehensive collection of biotoxin information from multiple perspectives. There are 14,607 data records with 8,975 toxins collected from 5,220 references and patents, covering more than 900 species. Due to the diversity of information on biotoxins, the distribution of data shows significant variability. Firstly, **Figure 1A** displays the top10 toxins in terms of the number of entries contributed to the database. Since there are many different types of toxins, it is challenging to use a single standard to measure and classify them. Therefore, we group toxins with similar physiological functions into the same category for statistical purposes to ensure the objectivity of the data. It can be seen that potassium channel toxins contribute the most data, with 1,046 entries, followed by sodium channel toxins with 744 entries. The number of entries corresponding to toxins that act on these two types of ion channels accounts for more than 10% of the total database, highlighting the significance of these toxins. Ranking tenth are acetylcholine receptor (nAChR) toxins, which only have 86 entries. These data differences indicate a significant disparity in the amount of information contained among different toxins. Similarly, **Figure 1B** shows the top10 targets in terms of the number of entries contributed to the database. The top two are still $K_v$ and $Na_v$, with 3,612 and 1,605 entries, respectively. Their number of entries far exceeds those of other targets, and their combined total accounts for more than 35% of the database. These data indicate that $Na_v$ and $K_v$ proteins remain popular targets for toxin research and drug design. Furthermore, we separately counted the biological species sources corresponding to the toxins and targets, as shown in **Figure 1C** and **Figure 1D**. Among the species sources of toxins, there are 293 toxins derived from haplopelma hainanum (spider), followed by lycosa singoriensis (spider) with 222 toxins.

In the top10 species sources of toxins, spiders and scorpions predominantly occupy the majority, highlighting their significant role in biotoxin research once again. Among the biological species sources of targets, homo sapiens (human) is the highest with 125 entries, followed by various rodents such as rattus norvegicus (82 entries) and mus musculus (48 entries). The tenth-ranked periplaneta americana has only 3 entries. These statistics on the biological species sources of targets indicate that research on mammals is much more extensive compared to other organisms.

In addition to extensive data collection, BioTD also has meticulously classified and processed the data information. BioTD has recorded 8,185 activity data records, encompassing a variety of activity data. **Figure 1E** shows the distribution of activity data, which can be mainly divided into five categories: Activity (such as effective concentration/dose, inhibition concentration), Hemolysis (such as hemolysis rate, minimum coagulant dose, minimum hemagglutination concentration), Kinetics (such as $K_d$, $K_i$, $K_{act}$, $Tau_{(off)}$), Safety (lethal concentration/dose, therapeutic index), and other physiological indicators. Among these, the data related to activity is the most numerous, reaching 4,454 entries, accounting for 54.4%. This is followed by other physiological indicators and kinetics data, with 2,512 entries (30.7%) and 1,129 entries (13.8%) respectively. Hemolysis activity and safety data are relatively few, with 48 entries (0.59%) and 42 entries (0.51%) respectively. In particular, BioTD has recorded the valuable 986 entries of biotoxin mutant data, accounting for 11.0% of all toxins, which provide crucial information for biotoxin research (**Figure 1F**). Furthermore, BioTD offers 3D visualization structures of all biotoxins to users. There are 1,753 (21.4%) experimentally resolved structures from PDB, and the remaining 7,222 (88.2%) structures are predicted using AlphaFold software (**Figure 1G**). These visualization structures will provide better user experience. Finally, BioTD has predicted and separated the toxin sequences containing the signal sequence in the database. As shown in **Figure 1H**, the number of toxins with signal sequence is 3,930, accounting for 43.8%, while the number of toxins without signal sequence is 5,045, accounting for 56.2%. These distinctions are beneficial for better understanding of the structure and function of biotoxins.

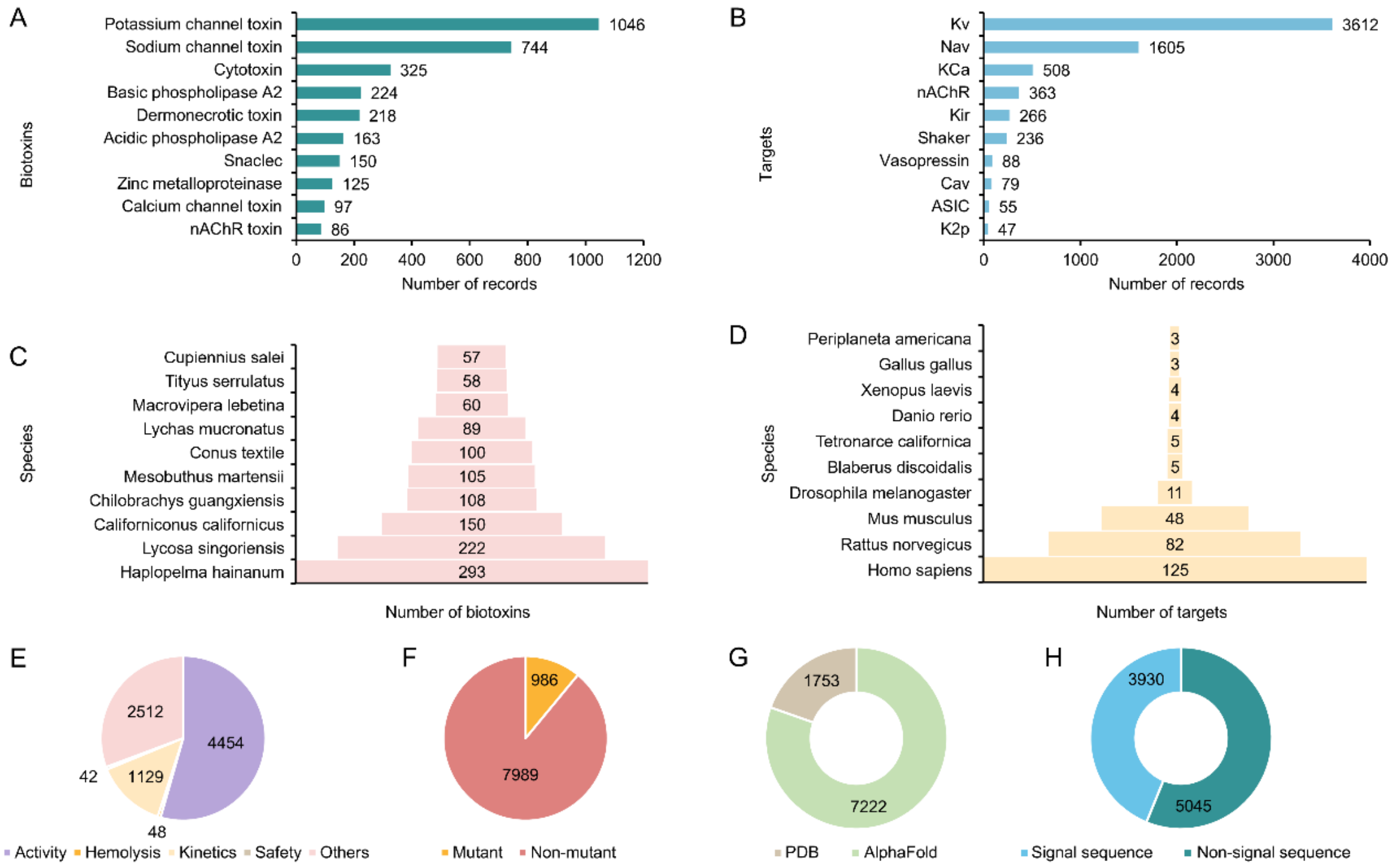

**Figure 1. Statistics in BioTD.** (A) and (B) The top10 biotoxins and targets in the database. nAChR: acetylcholine receptor; ASIC: acid-sensing ion channel; K2p: two pore-domain potassium channel. (C) and (D) The top10 species corresponding to the biotoxins and targets in the database. (E)-(H) The proportion of the activity data, biotoxin mutant, protein structure and signal sequence in the database.

## 3.2 Search of database

BioTD provides rich and fast search functions, which mainly realizes three types of searches: search for toxin, search for target and sequence similarity search (**Figure 2**). The main page search box can directly search for information related to toxins. Firstly, in the toxin search function, users can find toxin entries by searching for related name information, which includes the common name, Uniprot name and gene name among the entire textual component of BioTD. The resulting webpage displays profiles of all the toxins directly related to the search term, including the toxin name, toxin species, representative target, target family, target species and their information links. In order to facilitate a more customized input query, the wild characters of '*' and '?' are also supported. Users can obtain richer fuzzy search results through these wild characters. Furthermore, to enhance page interactivity, we provide the option-based search function below the main search box on the toxin search page. Users can first select the target name and then search for the corresponding toxins. Similarly, for target searches, users can search by target names, including the common name, Uniprot name, and gene name. The resulting webpage displays profiles of all the targets directly related to the search term, including target name, target bioclass, target species, representative toxin and their information links. The fuzzy search matching is also supported. On the interactive search page, we categorize targets biologically (including cells, coagulation factor, enzyme, immunoglobulin, receptor, transporter and channel). Users can first select the biological type of the target and then search for the corresponding target.

The toxins sequence similarity search function provides users with a more accurate and comfortable experience. On this page, the user can enter and submit the sequence in FASTA format of any protein, and BioTD will perform a sequence-based protein similarity search by using the BLASTp-based molecular sequence similarity search method. The similarity degree of identified molecules will be evaluated by the BLAST program, and with the BLAST E-value listed out in the order from the highest to the lowest. Links to the detailed information of identified toxins are also provided. On the search results page, BioTD provides a list of toxins with significant similarity to the sequence entered by the user, all molecules are sorted by E-value size, and the results table provides toxin ID, toxin name, sequence length, percentage of identity (%), BLAST score (bit) and E-value. The user can click on the Toxin ID to access the toxin details page.

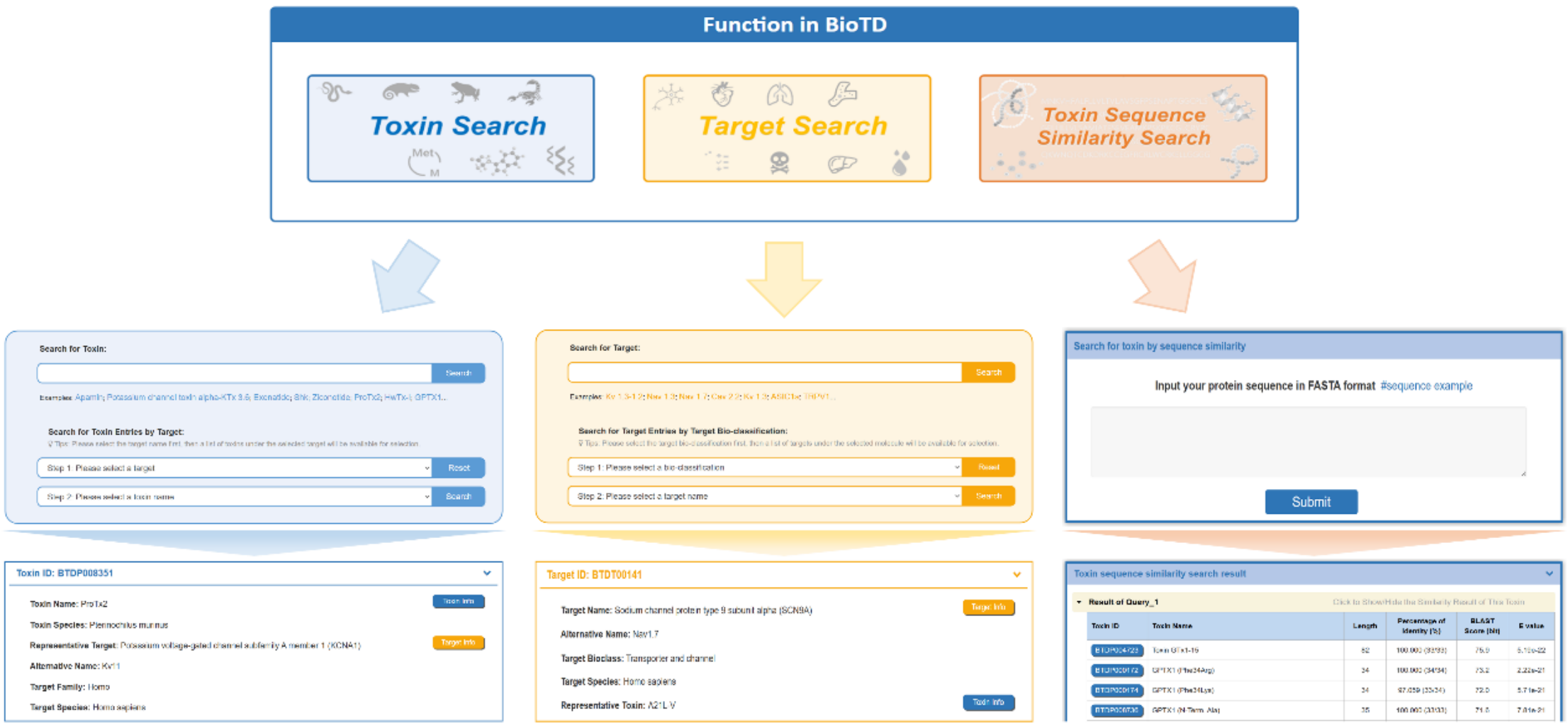

**Figure 2. The search function of BioTD.**

### 3.3 Biotoxin and target information

The ‘Biotoxin Page’ and ‘Target Page’ contain the main information of BioTD. In the ‘Biotoxin Page’, there are three parts: general information, activity data and references (**Figure 3**). The general information mainly contains the peptide ID, names, species, PDB/AlphaFold ID, sequence, mass, disulfide bond and species lineage of toxins. Users can click the Uniprot name to jump to the corresponding pages. All toxins in the database have a visual structure. The 3D stereoscopic structures of toxins are displayed in the window, and users can click the structure to rotate and zoom in/out. These toxin structures are derived from the PDB database and precited by AlphaFold software, and all structures can be downloaded freely in FASTA and PDB file format. Furthermore, we provide users with the full-length sequence of the toxin and the sequence after the removal of the signal peptide, and highlight the disulfide bond information, which will help users gain a deeper understanding of the toxin. All activity data of this toxin are listed in a table, which contains the target name, activity data type, activity data, concentration, note and reference. Users can click the ‘target info’ button to jump to the ‘Target Page’, and click the reference number to view the corresponding literature. All references are listed in order below and can be redirected to the PubMed page.

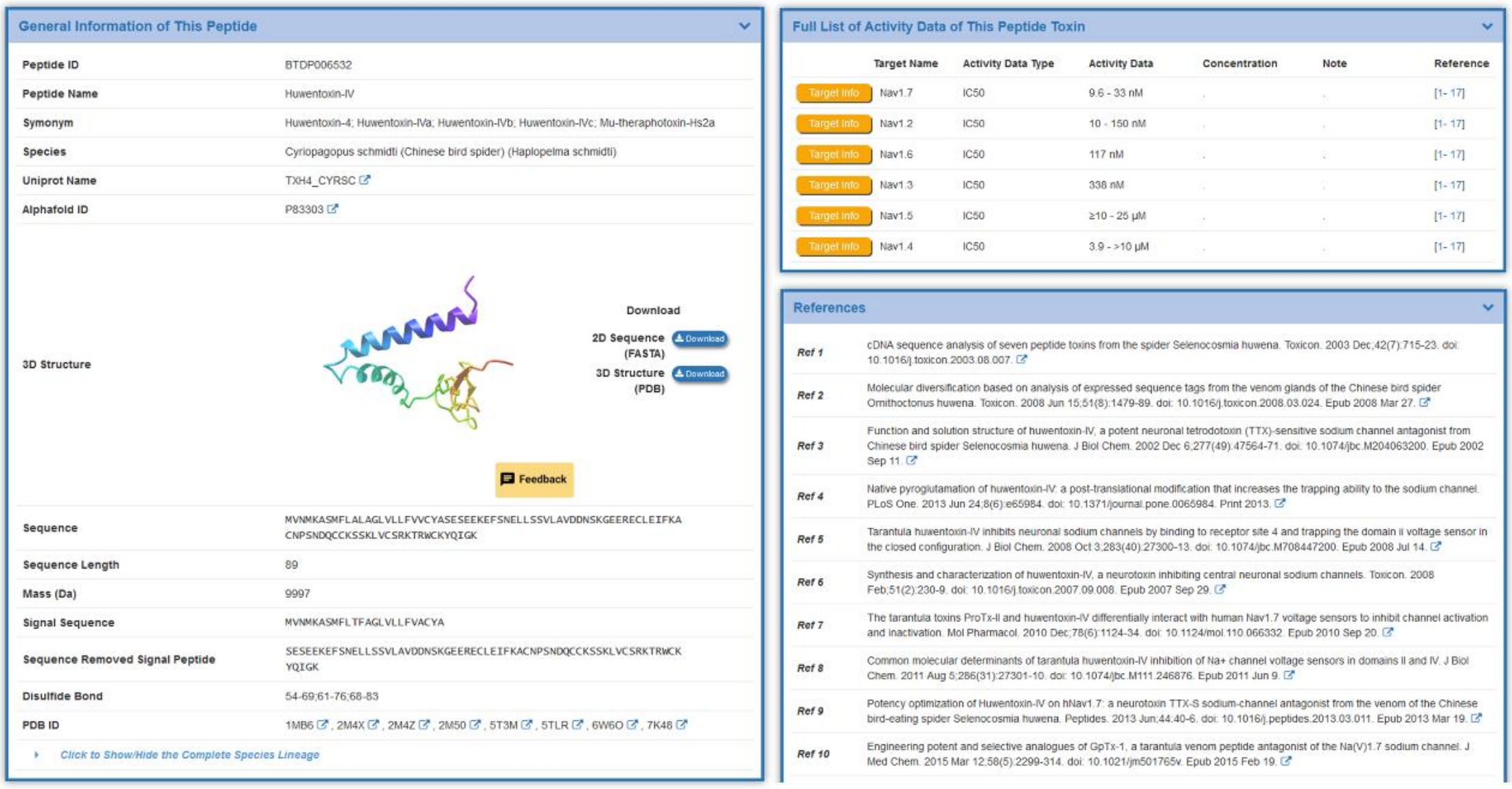

**Figure 3. The biotoxin page of BioTD.**

Similarly, in the 'Target Page', there are three parts: general information, toxin information and references (**Figure 4**). The general information mainly contains the target ID, names, target bioclass, sequence, family, function and species lineage of targets. Users can click the Uniprot ID, gene ID, Taxonomy ID and TCDB ID to jump to the corresponding pages. To facilitate user understanding of the target, we provide as many synonyms for the target name as possible and briefly describe the main functions of the target. All toxin information related to this target are listed in a table, which contains the toxin name, activity data type, activity data, and reference. Users can click the 'toxin info' button to jump to the 'Biotoxin Page', and click the reference number to view the corresponding literature.

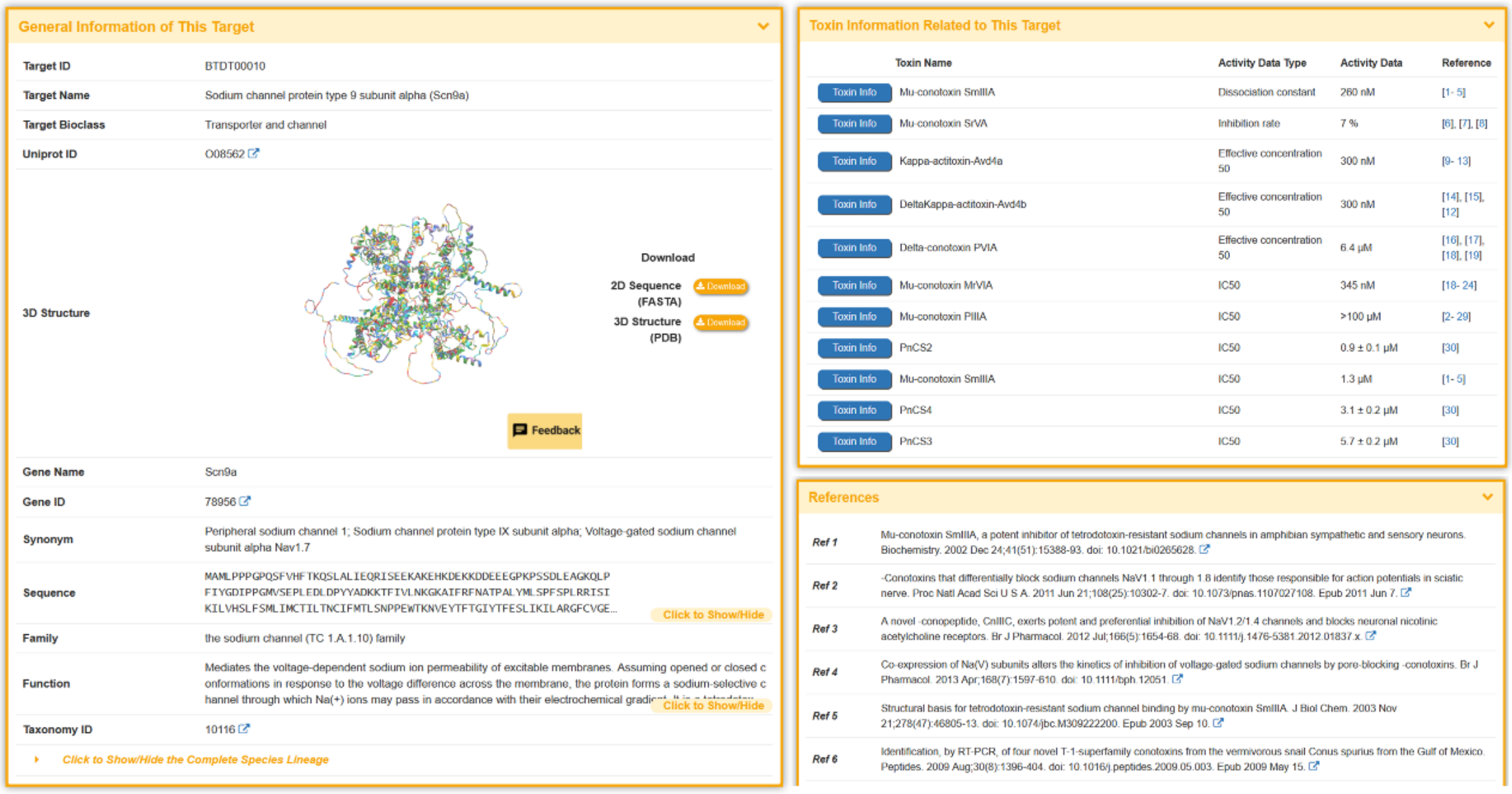

**Figure 4. The target page of BioTD.**

## 4 CONCLUSION

BioTD, the largest open-source database dedicated to biotoxins, aims to provide free data resource for toxins research and drug discovery is developed. We make great efforts to find and gather information for this database from all relevant public databases and publications. To make the data easier for users to access and use, we optimize the database in various ways. We hope that the BioTD can provide convenience for users, and we also expect users can give us more feedback to help us improve. All in all, we will keep mining the original data and expanding the database, and strive to develop some computational capabilities including molecular generation and virtual screening models. We believe that BioTD will serve as an important resource and a powerful tool for studying toxins and further drug design of venom-derived drugs.

## ACKNOWLEDGEMENTS

the National Natural Science Foundation of China (32371322, 32271329).

## REFERENCES

1. Smallwood, T. B.; Clark, R. J., Advances in venom peptide drug discovery: where are we at and where are we heading? *Expert Opin Drug Discov* **2021**, 16, 1163-1173.
2. UniProt, C., UniProt: the Universal Protein Knowledgebase in 2023. *Nucleic Acids Res* **2023**, 51, D523-D531.
3. Mendez, D.; Gaulton, A.; Bento, A. P.; Chambers, J.; De Veij, M.; Felix, E.; Magarinos, M. P.; Mosquera, J. F.; Mutowo, P.; Nowotka, M.; Gordillo-Maranon, M.; Hunter, F.; Junco, L.; Mugumbate, G.; Rodriguez-Lopez, M.; Atkinson, F.; Bosc, N.; Radoux, C. J.; Segura-Cabrera, A.; Hersey, A.; Leach, A. R., ChEMBL: towards direct deposition of bioassay data. *Nucleic Acids Res* **2019**, 47, D930-D940.
4. Burley, S. K.; Bhikadiya, C.; Bi, C.; Bittrich, S.; Chao, H.; Chen, L.; Craig, P. A.; Crichlow, G. V.; Dalenberg, K.; Duarte, J. M.; Dutta, S.; Fayazi, M.; Feng, Z.; Flatt, J. W.; Ganesan, S.; Ghosh, S.; Goodsell, D. S.; Green, R. K.; Guranovic, V.; Henry, J.; Hudson, B. P.; Khokhriakov, I.; Lawson, C. L.; Liang, Y.; Lowe, R.; Peisach, E.; Persikova, I.; Piehl, D. W.; Rose, Y.; Sali, A.; Segura, J.; Sekharan, M.; Shao, C.; Vallat, B.; Voigt, M.; Webb, B.; Westbrook, J. D.; Whetstone, S.; Young, J. Y.; Zalevsky, A.; Zardecki, C., RCSB Protein Data Bank (RCSB.org): delivery of experimentally-determined PDB structures alongside one million computed structure models of proteins from artificial intelligence/machine learning. *Nucleic Acids Res* **2023**, 51, D488-D508.
5. Wishart, D. S.; Feunang, Y. D.; Guo, A. C.; Lo, E. J.; Marcu, A.; Grant, J. R.; Sajed, T.; Johnson, D.; Li, C.; Sayeeda, Z.; Assempour, N.; Iynkkaran, I.; Liu, Y.; Maciejewski, A.; Gale, N.; Wilson, A.; Chin, L.; Cummings, R.; Le, D.; Pon, A.; Knox, C.; Wilson, M., DrugBank 5.0: a major update to the DrugBank database for 2018. *Nucleic Acids Res* **2018**, 46, D1074-D1082.
6. Krylov, N. A.; Tabakmakher, V. M.; Yureva, D. A.; Vassilevski, A. A.; Kuzmenkov, A. I., Kalium 3.0 is a comprehensive depository of natural, artificial, and labeled polypeptides acting on potassium channels. *Protein Sci* **2023**, 32, e4776.
7. Tunyasuvunakool, K.; Adler, J.; Wu, Z.; Green, T.; Zielinski, M.; Zidek, A.; Bridgland, A.; Cowie, A.; Meyer, C.; Laydon, A.; Velankar, S.; Kleywegt, G. J.; Bateman, A.; Evans, R.; Pritzel, A.; Figurnov, M.; Ronneberger, O.; Bates, R.; Kohl, S. A. A.; Potapenko, A.; Ballard, A. J.; Romera-Paredes, B.; Nikolov, S.; Jain, R.; Clancy, E.; Reiman, D.; Petersen, S.; Senior, A. W.; Kavukcuoglu, K.; Birney, E.; Kohli, P.; Jumper, J.; Hassabis, D., Highly accurate protein structure prediction for the human proteome. *Nature* **2021**, 596, 590-596.
8. Schoch, C. L.; Ciufo, S.; Domrachev, M.; Hotton, C. L.; Kannan, S.; Khovanskaya, R.; Leipe, D.; McVeigh, R.; O'Neill, K.; Robbertse, B.; Sharma, S.; Soussov, V.; Sullivan, J. P.; Sun, L.; Turner, S.; Karsch-Mizrachi, I., NCBI Taxonomy: a comprehensive update on curation, resources and tools. *Database (Oxford)* **2020**, 2020.
9. Lei, Z.; Shi Hong, L.; Li, L.; Tao, Y. G.; Yong Ling, W.; Senga, H.; Renchi, Y.; Zhong Chao, H., Batroxobin mobilizes circulating endothelial progenitor cells in patients with deep vein thrombosis. *Clin Appl Thromb Hemost* **2011**, 17, 75-9.
10. Warkentin, T. E.; Greinacher, A.; Koster, A., Bivalirudin. *Thromb Haemost* **2008**, 99, 830-9.
11. Lupi, A.; Rognoni, A.; Cavallino, C.; Secco, G. G.; Reale, D.; Cossa, G.; Rosso, R.; Bongo, A. S.; Cortese, B.; Angiolillo, D. J.; Jaffe, A. S.; Porto, I., Intracoronary vs intravenous bivalirudin bolus in ST-elevation myocardial infarction patients treated with primary angioplasty. *Eur Heart J Acute Cardiovasc Care* **2016**, 5, 487-96.
12. Cushman, D. W.; Ondetti, M. A., Design of angiotensin converting enzyme inhibitors. *Nat Med* **1999**, 5, 1110-3.

13. Warkentin, T. E., Bivalent direct thrombin inhibitors: hirudin and bivalirudin. *Best Pract Res Clin Haematol* **2004**, 17, 105-25.
14. Patchett, A. A., The chemistry of enalapril. *Br J Clin Pharmacol* **1984**, 18 Suppl 2, 201S-207S.
15. O'Shea, J. C.; Tcheng, J. E., Eptifibatide: a potent inhibitor of the platelet receptor integrin glycoprotein IIb/IIIa. *Expert Opin Pharmacother* **2002**, 3, 1199-210.
16. Furman, B. L., The development of Byetta (exenatide) from the venom of the Gila monster as an anti-diabetic agent. *Toxicon* **2012**, 59, 464-71.
17. Lee, C. J.; Ansell, J. E., Direct thrombin inhibitors. *Br J Clin Pharmacol* **2011**, 72, 581-92.
18. Hartman, G. D.; Egbertson, M. S.; Halczenko, W.; Laswell, W. L.; Duggan, M. E.; Smith, R. L.; Naylor, A. M.; Manno, P. D.; Lynch, R. J.; Zhang, G.; et al., Non-peptide fibrinogen receptor antagonists. 1. Discovery and design of exosite inhibitors. *J Med Chem* **1992**, 35, 4640-2.
19. Schmidtko, A.; Lotsch, J.; Freynhagen, R.; Geisslinger, G., Ziconotide for treatment of severe chronic pain. *Lancet* **2010**, 375, 1569-77.
20. Almagro Armenteros, J. J.; Tsirigos, K. D.; Sonderby, C. K.; Petersen, T. N.; Winther, O.; Brunak, S.; von Heijne, G.; Nielsen, H., SignalP 5.0 improves signal peptide predictions using deep neural networks. *Nat Biotechnol* **2019**, 37, 420-423.